%% file: main.tex
\documentclass[letter,11pt]{article}

\usepackage{setspace}
\usepackage{subcaption}
\usepackage{enumerate}
\usepackage{amsmath}
\usepackage{amsfonts}
\usepackage{graphicx}
\usepackage{tcolorbox}
\usepackage{amssymb}
\usepackage{multirow}
\usepackage{geometry}
\usepackage{soul}
\usepackage{colortbl}
\usepackage{wrapfig}
\usepackage{algorithm}
\usepackage{algorithmicx}
\usepackage{algpseudocode}
\usepackage{amsthm}
\usepackage{todonotes}
\usepackage{mathtools}
\usepackage{mathrsfs}
\usepackage{xspace}

\usepackage[pdftex, plainpages = false, pdfpagelabels, 
                 bookmarks=false,
                 bookmarksopen = true,
                 bookmarksnumbered = true,
                 breaklinks = true,
                 linktocpage,
                 pagebackref,
                 colorlinks = true,  % was true
                 linkcolor = blue,
                 urlcolor  = blue,
                 citecolor = red,
                 anchorcolor = green,
                 hyperindex = true,
                 hyperfigures
                 ]{hyperref}

\usepackage{thmtools} 
\usepackage{thm-restate}
\usepackage{bm}

\algtext*{EndWhile}% Remove "end while" text
\algtext*{EndIf}% Remove "end if" text
\algtext*{EndFor}% Remove "end for" text

\newtheorem{lemma}{Lemma}

\newtheorem{theorem}[lemma]{Theorem}
\newtheorem{theorem*}[lemma]{Theorem*}

\newtheorem{observation}[lemma]{Observation}

\newcommand{\ignore}[1]{}
\newcommand{\R}{\mathbb{R}}
\newcommand{\opt}{\textrm{opt}}

\newcommand{\eps}{\varepsilon}

\newcommand{\wind}{\mathsf{wind}}

\newcommand{\EnclosingPoints}{\textsc{Enclosing-All-Points}\xspace}

\title{\LARGE Enclosing Points with Geometric Objects}
\author{Timothy M. Chan\footnote{University of Illinois at Urbana-Champaign, USA. \texttt{tmc@illinois.edu}.} \and Qizheng He\footnote{University of Illinois at Urbana-Champaign, USA. \texttt{qizheng6@illinois.edu}.} \and Jie Xue\footnote{New York University Shanghai, China. \texttt{jiexue@nyu.edu}.}}
\date{}

\begin{document}

\maketitle

\bibliographystyle{plainurl}

\begin{abstract}
%We study the problem of enclosing a set of points in the plane using fewest geometric objects as obstacles.
Let $X$ be a set of points in $\mathbb{R}^2$ and $\mathcal{O}$ be a set of geometric objects in $\mathbb{R}^2$, where $|X| + |\mathcal{O}| = n$.
We study the problem of computing a minimum subset $\mathcal{O}^* \subseteq \mathcal{O}$ that encloses all points in $X$.
Here a point $x \in X$ is enclosed by $\mathcal{O}^*$ if it lies in a bounded connected component of $\mathbb{R}^2 \backslash (\bigcup_{O \in \mathcal{O}^*} O)$.
We propose two algorithmic frameworks to design polynomial-time approximation algorithms for the problem.
The first framework is based on sparsification and min-cut, which results in $O(1)$-approximation algorithms for unit disks, unit squares, etc.
The second framework is based on LP rounding, which results in an $O(\alpha(n)\log n)$-approximation algorithm for segments, where $\alpha(n)$ is the inverse Ackermann function, and an $O(\log n)$-approximation algorithm for disks.
\end{abstract}

%\tableofcontents
%\newpage
%\input{intro}

\input{introduction}

\input{pre}
\input{Sparseframe}
\input{LPframe}

%\input{conclusion}

%\paragraph{Acknowledgement.} $\sharp\sharp\sharp$
%\bibliographystyle{plain}
\bibliography{references}
%\appendix
%\input{appen}

\appendix

\section{More results based on sparsification and min-cut} \label{app-framework2}
Our algorithmic technique in Section~\ref{sec-sparsecut} can be generalized to any set $\mathcal{O}$ of obstacles that are similarly-sized fat pseudo-disks.
Formally, these obstacles are required to satisfy the following conditions.
First, all obstacles are convex, and there exist constants $r^-$ and $r^+$ such that for every $O \in \mathcal{O}$, there exist two concentric disks $D_O^-$ and $D_O^+$ with radii $r^-$ and $r^+$ respectively satisfying that $D_O^- \subseteq O \subseteq D_O^+$.
Second, the boundaries of any two obstacles intersect at most twice.

To generalize our algorithm, we need to choose the center $\mathsf{ctr}(O)$ of each obstacle $O \in \mathcal{O}$ and specify a curve $\sigma(O,O')$ for every edge $\mathsf{Int}(\mathcal{O})$.
By definition, for every $O \in \mathcal{O}$, we have concentric disks $D_O^-$ and $D_O^+$ with radii $r^-$ and $r^+$ respectively satisfying $D_O^- \subseteq O \subseteq D_O^+$.
We simply define $\mathsf{ctr}(O)$ as the common center of $D_O^-$ and $D_O^+$.
To define $\sigma(O,O')$, we choose an arbitrary point $o \in O \cap O'$.
Then we let $\sigma(O,O')$ be the poly-line obtained by concatenating the segment between $\mathsf{ctr}(O)$ and $o$ with the segment between $o$ and $\mathsf{ctr}(O')$.
Clearly, $\sigma(O,O') \subseteq O \cup O'$.

The arguments are almost the same as those in Section~\ref{sec-sparsecut}.
The only place we need to slightly adjust is the argument for Observation~\ref{obs-twopairs}.
Consider a relevant pair $(\Box,\Box')$, and let $Y \subseteq X$ consist of the points enclosed by $\mathcal{O}_\Box \cup \mathcal{O}_{\Box'}$ where $\mathcal{O}_\Box = \{O \in \mathcal{O}: \mathsf{ctr}(O) \in \Box\}$ and $\mathcal{O}_{\Box'} = \{O \in \mathcal{O}: \mathsf{ctr}(O) \in \Box\}$.
As in Section~\ref{sec-sparsecut}, let $\mathsf{Int}^* \subseteq \mathsf{Int}(\mathcal{O}_\Box \cup \mathcal{O}_{\Box'})$ consist of the edges $(O,O')$ where $O \in \mathcal{O}_\Box$ and $O' \in \mathcal{O}_{\Box'}$.
Here we need a definition for $Y_L(O,O')$ and $Y_R(O,O')$ for every $(O,O') \in \mathsf{Int}^*$ so that the property in Observation~\ref{obs-twopairs}, which can in turn allow us to use the same argument to show the existence of four objects in $\mathcal{O}_\Box \cup \mathcal{O}_{\Box'}$ enclosing $Y$.
Denote by $z$ and $z'$ the centers of $\Box$ and $\Box'$, respectively.
For convenience, we move $\mathsf{ctr}(O)$ for all $O \in \mathcal{O}_\Box$ (resp., $O \in \mathcal{O}_{\Box'}$) to the point $z$ (resp., $z'$).
This is fine because we still have $\mathsf{ctr}(O) \in \Box$ for all $O \in \mathcal{O}_\Box$ and $\mathsf{ctr}(O) \in \Box'$ for all $O \in \mathcal{O}_{\Box'}$.
Now for every $(O,O') \in \mathsf{Int}^*$, $\sigma(O,O')$ is a two-piece poly-line connecting $z$ and $z'$.
Let $\ell$ be the line through $z$ and $z'$.
For $(O,O') \in \mathsf{Int}^*$, let $\ell(O,O')$ be the poly-line obtained from $\ell$ by replacing the segment $\overline{zz'}$ with $\sigma(O,O')$.
Then $\ell(O,O')$ partitions the plane into two parts $H_L$ and $H_R$, where $H_L$ (resp., $H_R$) is to the left (resp., right) of $\ell(O,O')$ with respect to the direction from $z$ to $z'$.
Now we can define $Y_L(O,O') = Y \cap H_L$ and $Y_R(O,O') = Y \cap H_R$.

To see the property in Observation~\ref{obs-twopairs} holds, consider $(O,O'),(P,P') \in \mathsf{Int}^*$.
If $\sigma(O,O')$ and $\sigma(P,P')$ do not intersect at any point other than $z$ and $z'$, the same proof of Observation~\ref{obs-twopairs} works.
So assume $\sigma(O,O')$ and $\sigma(P,P')$ intersect at a point $v$ other than $z$ and $z'$.
%Without loss of generality, suppose $v \in O$ and $v \in P'$.
Let $o \in O \cap O'$ (resp., $p \in P \cap P'$) be the middle vertex of $\sigma(O,O')$ (resp., $\sigma(P,P')$).
Without loss of generality, assume $\overline{zo}$ and $\overline{z'p}$ intersect at the point $v$.
Then $v \in O$ and $v \in P'$.
As in the proof of Observation~\ref{obs-twopairs}, our goal is to show that either $Y \cap \triangle zvp = \emptyset$ or $Y \cap \triangle z'vo = \emptyset$, which implies either $Y_L(O,O') \subseteq Y_L(P,P')$ or $Y_L(P,P') \subseteq Y_L(O,O')$.
If $p \in O$, then $z,v,p \in O$ and thus $\triangle zvp \subseteq O$, which implies $Y \cap \triangle zvp = \emptyset$.
Also, if $z' \in O$, then $z',v,o \in O$ and thus $\triangle z'vo \subseteq O$, which implies $Y \cap \triangle z'vo = \emptyset$.
So assume $p,z' \notin O$.
For the same reason, we can assume $o,z \notin P'$.
But this contradicts the pseudo-disk property of $O$ and $P'$.
To see this, let $\ell_1$ (resp., $\ell_2$) be the line through $o$ and $z$ (resp., $p$ and $z'$).
The lines $\ell_1$ and $\ell_2$ intersect at $v$ (which is contained in both $O$ and $P'$), and partition the plane into four wedges.
Since the boundaries of $O$ and $P'$ intersect at most twice, there exists one wedge $W$ in which the boundaries of $O$ and $P'$ do not intersect and hence either $O \cap W \subseteq P' \subseteq W$ or $P' \cap W \subseteq O \subseteq W$.
However, observe that every wedge contains a point in $O \backslash P'$ and a point in $P' \backslash O$, and hence cannot satisfy this condition.
For example, the wedge whose boundary containing $\overline{zv}$ and $\overline{pv}$ contains $z \in O \backslash P'$ and $p \in P' \backslash O$.
Therefore, we must have either $Y \cap \triangle zvp = \emptyset$ or $Y \cap \triangle z'vo = \emptyset$, and everything follows.

\begin{theorem}\label{thm:enclosing_segments}
There exists a polynomial-time $O(1)$-approximation algorithm for \EnclosingPoints with similarly-sized fat pseudo-disks (in particular, unit disks and unit squares), where $n$ is the total number of points and obstacles.
\end{theorem}

\section{More results based on LP} \label{app-framework1}
The LP-based technique can be modified to work for more general types of curves with minor changes.
For curves that pairwise intersect only $s=O(1)$ times, one technical issue is during unwinding, a simple cycle may only have length $2$.
If we decompose a cycle $C$ into a cycle $C_1$ with length $2$ and another cycle $C_2$, then the length of $C_2$ is the same as $C$, therefore not getting a good recurrence for the maximum total size $f(n)$ of the simple cycles that we decompose into.

To fix this issue, we first try to unwind at a self-intersection point $v$ of $C$, such that the two subcycles have length strictly less than $\mathsf{len}(C)$. If such point exists, then our previous analysis for the recurrence of $f(n)$ still works.
Otherwise if there are no such points, then it means only pairs of consecutive edges on the cycle $C$ can intersect.
In this case, there are only at most $s\cdot \mathsf{len}(C)$ self-intersections on $C$, so we can decompose $C$ into $O(s \cdot \mathsf{len}(C))$ subcycles, increasing the total number of edges by only a factor of $O(s)=O(1)$.
%otherwise can unwind at another intersection point.

%Let $f(n)$ denote the maximum total number of edges that a non-simple cycle of length $n$ can decompose into. $f(n)$ satisfies the recurrence
%\[f(n)\leq \max_{2\leq n_1\leq n} \left(f(n_1)+f(n-n_1+2)\right),\]

For curves that pairwise intersect at most $s$ times, it is known that the combinatorial complexity of the outer face in the arrangement is bounded by $O(\lambda_{s+2}(n))$~\cite{DBLP:journals/dcg/GuibasSS89,AgarwalS00} as mentioned earlier in Sec.~\ref{sec:enclosing_preliminaries}. So our approximation algorithm yields the following result.
%curved segments
%(or objects with linear union complexity)
%(or polygons with constant complexity)
\begin{theorem}\label{thm:enclosing_curves}
There exists a polynomial-time $O(\frac{\lambda_{s+2}(n)}{n} \log n)$-approximation algorithm for \EnclosingPoints with curves that pairwise intersect at most $s=O(1)$ times, where $n$ is the total number of points and segments.
%Given a set $X$ of points and a set $S$ of colored curves that pairwise intersect at most $s=O(1)$ times in $\R^2$, where $|X|+|S|=n$, there exists a polynomial-time $O(\frac{\lambda_{s+2}(n)}{n}\log n)$-approximation algorithm for enclosing all points in $X$ using the smallest subset of objects in $S$.
\end{theorem}

In particular, the approach also applies to disks.
Since it is known that the outer boundary for disks has complexity $O(n)$, we get the following result for disks.
\begin{theorem}\label{thm:enclosing_disks}
There exists a polynomial-time $O(\log n)$-approximation algorithm for \EnclosingPoints with disks, where $n$ is the total number of points and disks.
%Given a set $X$ of points and a set $S$ of colored disks in $\R^2$, where $|X|+|S|=n$, there exists a polynomial-time $O(\log n)$-approximation algorithm for enclosing all points in $X$ using the smallest subset of objects in $S$.
\end{theorem}

\end{document}

%% file: introduction.tex
\section{Introduction}

Studying problems related to plane obstacles is a popular topic in computational geometry.
In the common setting of such problems, we are given a set of geometric objects in the plane as \textit{obstacles}.
An obstacle blocks all paths in the plane it intersects.
Various optimization problems have been investigated in this setting.
For example, a line of research~\cite{asano1986visibility,ghosh1991output,hershberger1999optimal,kapoor1988efficient,mitchell1993shortest,wang2023new} focused on finding shortest paths between two points admidst the given obstacles.
The obstacle-removal problem~\cite{ChanK12,KumarLSS21,0004L0S022,KumarLA07} asks for a minimum subset of obstacles such that after removing these obstacles, two specified points in the plane have a path between them.
The point-separation problem~\cite{CabelloG16,GibsonKPVV16,0004L0S022} aims to select a minimum subset of obstacles that separate a given set $X$ of points, i.e., block all paths between two points in $X$.

\begin{figure}[t]\centering
    \includegraphics[width=.7\textwidth]{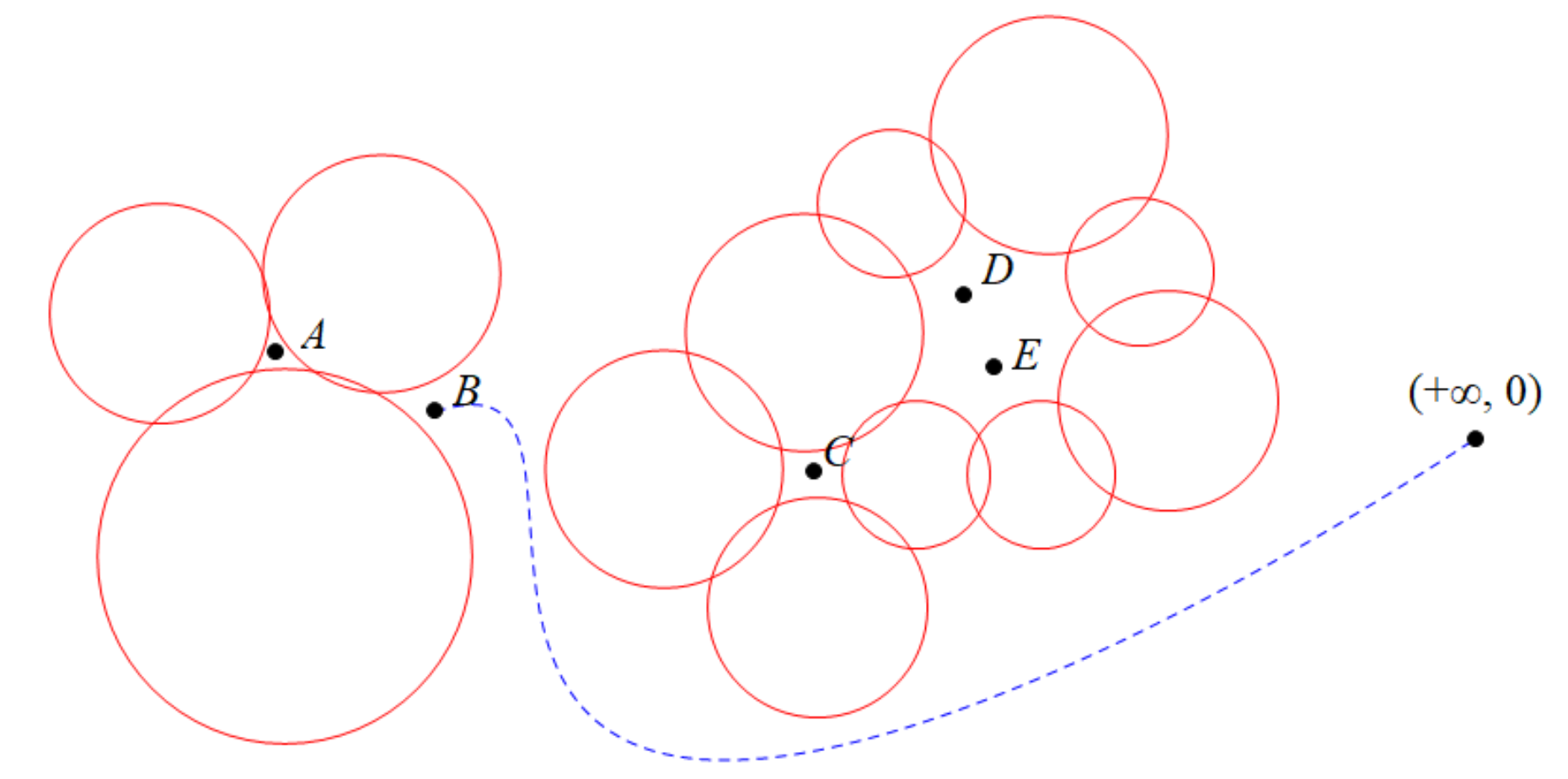}\\
    \caption{The set of red disks enclose points $A$, $C$, $D$ and $E$, but does not enclose the point $B$.}
    \label{fig:enclosing_points}
\end{figure}

%After studying the geometric set cover problem in the previous parts, we further consider a problem that is closely related to geometric set cover: enclosing all points with geometric objects (\EnclosingPoints). The problem is defined as follows:
In this paper, we study a natural problem related to plane obstacles, which we call \EnclosingPoints.
We say a point $x$ in the plane is \textit{enclosed} by a set $\mathcal{O}$ of (compact) geometric objects if $x$ lies in a bounded connected component of $\mathbb{R}^2 \backslash (\bigcup_{O \in \mathcal{O}} O)$; in other words, any curve connecting $x$ with the point $(+\infty,0)$ at infinity must intersect with at least one object in $\mathcal{O}$.
An example is shown in Figure~\ref{fig:enclosing_points}.
The problem simply aims to compute a minimum subset of obstacles to enclose all input points.
%This problem turns out to be related to geometric set cover, as we will see later.

\begin{tcolorbox}[colback=gray!5!white,colframe=gray!75!black]
\EnclosingPoints \\
\textbf{Input:} A set $X$ of points and a set $\mathcal{O}$ of geometric objects in $\R^2$. \\
\textbf{Output:} A minimum subset $\mathcal{O}^*\subseteq \mathcal{O}$ that enclose all points in $X$.
\end{tcolorbox}

\EnclosingPoints is closely related to the aforementioned point-separation problem.
In fact, it is a special case of the \textit{generalized} point-separation problem studied in~\cite{0004L0S022}, in which we want to use a minimum number of obstacles to separate a set of point-pairs $\{(s_1,t_1),\dots,(s_r,t_r)\}$ in the plane, i.e., block all paths between $s_i$ and $t_i$ for all $i \in [r]$.
Enclosing a point-set $X$ is equivalent to separating the point-pairs in $\{(x,z): x \in X\}$, where $z = (+\infty,0)$ is the point at infinity.
Kumar et al.~\cite{0004L0S022} showed that when the number of point-pairs to be separated is fixed, the generalized point-separation problem is polynomial-time solvable for any connected obstacles.
However, this does not solve the \EnclosingPoints problem, since in our problem the set $X$ is a part of the input (when the number of point-pairs is not fixed, no nontrivial algorithms for the generalized point-separation problem were known).
To the best of our knowledge, \EnclosingPoints has not been studied before.

In this paper, we propose two general algorithmic frameworks to design polynomial-time approximation algorithms for \EnclosingPoints.

\paragraph{Approach based on sparsification and min-cut.}
%Our second framework is less general than the first one in terms of obstacles, but it can give $O(1)$-approximation algorithms.
Our first framework solves \EnclosingPoints via two steps.
The first step computes a \textit{sparse} subset $\mathcal{O}' \subseteq \mathcal{O}$ of obstacles such that an optimal solution using the obstacles in $\mathcal{O}'$ is an $O(1)$-approximation solution of the original problem for $\mathcal{O}$.
Roughly speaking, here ``sparse'' means that any point in the plane stabs $O(1)$ obstacles in $\mathcal{O}'$.
The second step reduces the problem on $\mathcal{O}'$ to a min-cut problem in a graph (which is polynomial-time solvable) by losing a constant factor in cost.
This framework results in $O(1)$-approximation algorithms for the problem with similarly-sized fat pseudo-disks, e.g., unit-disks, unit squares, etc.

\paragraph{Approach based on LP rounding.}
Our second framework is based on LP rounding.
Compared to the first one, this framework applies to more general types of obstacles, while giving worse approximation ratios.
Specifically, it results in an $O(\alpha(n) \log n)$-approximation algorithm for \EnclosingPoints with line segments, where $\alpha(n)$ denotes the inverse Ackermann function, and an $O(\log n)$-approximation algorithm for \EnclosingPoints with (general) disks.
The framework also generalizes to the problem with curves that pairwise intersect $s = O(1)$ times, with an approximation ratio $\alpha(n)^{O(\alpha(n)^{s-1})} \log n$.
Although LP rounding is a common technique widely used for designing approximation algorithms, how to apply it for \EnclosingPoints is totally non-obvious.
In fact, for our problem, new ideas are required in both the LP formulation and the rounding scheme.

%\begin{itemize}
    %\item \textbf{Approach based on LP rounding.}
    %\item \textbf{Approach based on sparsification and min-cut.}
%\end{itemize}

\subsection{Related work}

%We remark that the problem can also be more generally defined in higher dimensions; however, all previous related works only study the problem in 2D (since the initial motivation is to isolate points by wireless sensors, which are 2D disks or unit disks). Designing approximation algorithms for this problem in dimension three or higher appears to be much harder. Therefore, in this thesis we also only focus on the 2D case.

%We start with surveying a number of similar problems related to the \EnclosingPoints problem we study here.

To the best of our knowledge, \EnclosingPoints has not been studied before.
Here we briefly summarize the literature for two problems that are closely related to \EnclosingPoints, the point-separation problem and the obstacle-removal problem.

%\paragraph{Related works.}
\paragraph{Point separation.}
Given a set $X$ of points and a set $\mathcal{O}$ of obstacles in $\R^2$, the point-separation problem asks for a minimum subset $\mathcal{O}^* \subseteq \mathcal{O}$ that separates all points in $X$.
%Given a set $S$ of geometric objects and a set $X$ of points in $\R^2$, the \PointsSeparation problem asks to select the minimum number of input objects, such that all pairs of points in $X$ are separated by the selected objects. We say a pair of points $p$ and $q$ are separated by a set $S^*$ of objects, if any curve connecting $p$ and $q$ intersect at least one object in $S^*$. As a comparison, our enclosing problem asks to separate all points with a single point $(+\infty,0)$ (the infinity), while the \PointsSeparation problem asks to separate all pairs of points.
This problem has applications in barrier coverage with wireless sensors.
%In real world, people use wireless sensors to guard or monitor buildings, estates, national borders etc, and the coverage region of the wireless sensors used can usually be modeled as unit disks or disks. If two points $p$ and $q$ are separated, then no intruder can get from one point to the other without getting noticed. Our enclosing problem is also naturally relevant to these applications, and this is one of the motivations for us to study this particular problem. Instead of covering the whole area with sensors, it is more economical to only monitor a boundary of the buildings, while still ensuring that the interior is guarded.
Gibson et al.~\cite{GibsonKPVV16} showed the NP-hardness of the problem for unit disks by reducing from planar multiterminal-cut, and presented an $(9+\varepsilon)$-approximation algorithm that works for (general) disks.
Later, Cabello and Giannopoulos~\cite{CabelloG16} designed the first polynomial-time exact algorithm for point-separation with arbitrary connected curves for the case $|X| = 2$, which runs in $O(n^3)$ time.
%with $O(n^3)$ running time is given for separating two points using arbitrary connected curves, assuming the curves have reasonable computational properties. (An earlier version of this paper appeared in a manuscript by Alt et al.\ \cite{alt2011minimum}, which only works for line segments.)
%They utilized a concept called the ``3-path-condition'' \cite{DBLP:journals/jct/Thomassen90b} (see also \cite[Chapter 4]{DBLP:books/daglib/0030489}), connecting the problem to topology.
They also showed NP-hardness of the problem with unit circles or orthogonal segments, via a reduction from planar-3-SAT.
Recently, Kumar et al.~\cite{0004L0S022} generalizes the algorithm of Cabello and Giannopoulos~\cite{CabelloG16}
and showed that point-separation with arbitrary connected curves can be solved in $n^{O(|X|)}$ time.

\paragraph{Obstacle removal.} 
Given two points $s,t$ and a set $\mathcal{O}$ of obstacles in $\R^2$, the obstacle problem asks for a minimum subset $\mathcal{O}^* \subseteq \mathcal{O}$ such that $s$ and $t$ are not separated by $\mathcal{O} \backslash \mathcal{O}^*$.
This problem is also sometimes called computing the \emph{barrier resilience} \cite{ChanK12,KumarLA07}, where a real-world application is to compute the minimum number of sensors that need to be deactivated, so that there is a path between $s$ and $t$ that does not intersect any of the active sensors.
It also has applications in robotics \cite{EibenGKY18,EricksonL13}.
%A related concept is called the \emph{thickness} of the barrier, counting the minimum number of intersections with the sensors over all paths from $p$ to $q$ (a path may intersect a sensor several times). Easy to see the resilience is always a lower bound on the thickness.
Obstacle-removal was shown to be NP-hard for arbitrary line segments \cite{alt2017minimum}, for unit segments \cite{tseng2011resilience,TsengK11}, and for certain types of fat regions with bounded ply (such as axis-aligned rectangles of aspect ratio $1:1+\eps$ and $1+\eps:1$) \cite{korman2018complexity}.
For approximation results, Bereg and Kirkpatrick provided a $3$-approximation algorithm for unit disks \cite{BeregK09}.
%In particular, they showed any (Euclidean) shortest path from $s$ to $t$ that intersects a fixed number of distinct sensors, never intersects the same sensor more than three times. As a corollary, the thickness is at most three times the resilience. On the other hand, it is not known whether for unit disks the problem is still NP-hard.
Bandyapadhyay et al.~\cite{BandyapadhyayKS20} presented an $O(\sqrt{n})$-approximation algorithm for pseudodisks and rectilinear polygons.
Kumar et al.~\cite{KumarLSS21} further designed an $O(1)$-approximation algorithm for any well-behaved objects, such as polygons or splines.
Their arguments were later simplified by Kumar et al.~\cite{0004L0S022}.

%% file: pre.tex
\section{Preliminaries}\label{sec:enclosing_preliminaries}

\paragraph{Curves and homotopy.}
A \textit{curve} in the plane is a continuous map $\gamma:[0,1] \rightarrow \mathbb{R}^2$.
For convenience, sometimes the term ``curve'' also refers to the image of such a map.
If $\gamma(0) = \gamma(1)$, we say $\gamma$ is a \textit{closed curve} and define its \textit{base point} as $\gamma(0) = \gamma(1)$.
A segment between two points $a$ and $b$ naturally defines two curves, one from $a$ to $b$ and the other from $b$ to $a$.
For two curves $\gamma,\gamma':[0,1] \rightarrow \mathbb{R}^2$ with $\gamma(1) = \gamma'(0)$, we can concatenate them to obtain a curve from $\gamma(0)$ to $\gamma'(1)$.

Two curves $\gamma,\gamma':[0,1] \rightarrow \mathbb{R}^2$ with $\gamma(0) = \gamma'(0)$ and $\gamma(1) = \gamma'(1)$ are \textit{homotopic} in a region $R \subseteq \mathbb{R}^2$ if the images of $\gamma,\gamma'$ lie in $R$ and one can deform $\gamma$ continuously to $\gamma'$ inside $R$ without changing the two endpoints $\gamma(0)$ and $\gamma(1)$.
A closed curve $\gamma$ is \textit{contractible} in $R \subseteq \mathbb{R}^2$ if $\gamma$ and the trivial curve $\mu$ are homotopic in $R$, where $\mu(i) = \gamma(0) = \gamma(1)$ for all $i \in [0,1]$; and \textit{non-contractible} in $R$ otherwise.

\paragraph{Union complexity of geometric objects.} For a type of geometric objects, its \emph{union complexity} is defined as the combinatorial complexity of the boundary of the union of $n$ such objects. It is known that disks have linear union complexity \cite{KedemLPS86}.

For our applications, it suffices to look at the combinatorial complexity of the outer face in the arrangement of the objects, which is upper-bounded by the union complexity of the objects. So for disks, this complexity is $O(n)$. Pollack et al.~\cite{PollackSS88} proved that for $n$ arbitrary line segments in $\R^2$, the complexity of the outer face in the arrangement is bounded by $O(n\alpha(n))$ (by Davenport-Schinzel sequences), where $\alpha(n)$ is the inverse Ackermann function.

More generally, for curves that pairwise intersect only at most $s=O(1)$ times, a similar bound holds: the combinatorial complexity of the outer face in the arrangement is $O(\lambda_{s+2}(n))$, where the function $\lambda_s(n)=n\alpha(n)^{O(\alpha(n)^{s-3})}$ is the maximum length of an $(n,s)$ Davenport-Schinzel sequence~\cite{AgarwalS00,DBLP:journals/dcg/GuibasSS89}. For any constant $s$, $\lambda_s(n)$ is almost linear in $n$.

\paragraph{Point-in-polygon check.} Given a point $q$ and a polygon $P$, the point-in-polygon problem asks whether $q$ lies inside, outside, or on the boundary of $P$. A standard method for solving this problem is to use the ray-casting algorithm \cite{Shimrat62}: cast a ray $\vec{r}$ starting from $q$ and going in any fixed direction, if $\vec{r}$ intersects $P$ an even number of times, then $q$ is outside $P$; otherwise if $\vec{r}$ intersects $P$ an odd number of times, then $q$ is inside $P$. (For our applications, we consider the points on the boundary of $P$ as inside $P$.)

\begin{figure}[!htbp]
    \centering
    \includegraphics[width=.55\textwidth]{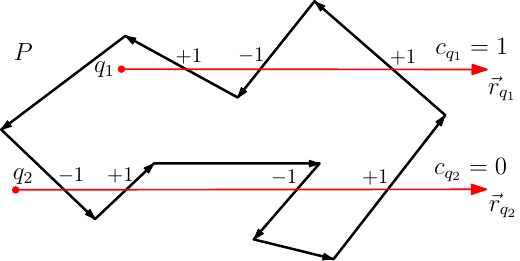}
    \caption{The winding numbers with respect to a (simple) polygon $P$. The point $q_1$ has winding number $c_{q_1}=1$, and is inside $P$; the point $q_2$ has winding number $c_{q_2}=0$, and is outside $P$.}
    \label{fig:winding_number}
\end{figure}

Another method for point-in-polygon check is by computing the \emph{winding number} of $q$ with respect to $P$, denote as $\wind(q,P)$. The idea is to modify the ray-casting algorithm as follows: first orient the edges of $P$ in counter-clockwise direction. Initialize a counter $c_q$ with value $0$. Cast a ray $\vec{r}_q$ starting from $q$ and going in any fixed direction, for each edge $\vec{e}$ of $P$ intersecting $\vec{r}_q$, if $\vec{e}$ crosses $\vec{r}_q$ in counter-clockwise direction, then increase $c_q$ by 1; otherwise if $\vec{e}$ crosses $\vec{r}_q$ in clockwise direction, decrease $c_q$ by 1. In the end, $q$ is inside $P$ iff the counter $c_q$ is nonzero. In particular, when $P$ is a simple polygon, $q$ is inside $P$ iff $c_q=1$. Intuitively, if $q$ is outside $P$, then the weighted crossings between the ray $\vec{r}_q$ and the polygon $P$ will cancel each other. An example for this process is shown in Fig.~\ref{fig:winding_number}.

%% file: sparseframe.tex
\section{Approach based on sparsification and min-cut} \label{sec-sparsecut}

In this section, we discuss our algorithmic framework based on sparsification and min-cut.
We present our algorithm for unit disks.
It generalizes to any similarly-sized fat pseudo-disks, e.g., unit squares or more generally translates of a fixed convex body; see Appendix~\ref{app-framework2}.

%Suppose $S = \{A_1,\dots,A_n\}$.
%We use $G$ to denote the intersection graphs of the unit disks in $\mathcal{D}$.
Let $X$ be a set of points in $\mathbb{R}^2$ and $\mathcal{D}$ be a set of unit disks.
For a subset $\mathcal{D}' \subseteq \mathcal{D}$, denote by $G[\mathcal{D}']$ the intersection graphs of the unit disks in $\mathcal{D}'$ and by $\mathsf{Int}(\mathcal{D}')$ the edge set of $G[\mathcal{D}']$, i.e., the set of pairs $(D,D')$ of unit disks in $\mathcal{D}'$ that intersect each other.
For each unit disk $D \in \mathcal{D}$, let $\mathsf{ctr}(D)$ denote the center of $D$.
For every edge $(D,D') \in \mathsf{Int}(\mathcal{D})$, we have $D \cap D' \neq \emptyset$ and let $\sigma(D,D')$ be the curve from $\mathsf{ctr}(D)$ to $\mathsf{ctr}(D')$ defined by the segment between $\mathsf{ctr}(D)$ and $\mathsf{ctr}(D')$; note that $\sigma(D,D') \subseteq D \cup D'$.
With a bit abuse of notation, we also use $\sigma(D,D')$ to denote the segment between $\mathsf{ctr}(D)$ and $\mathsf{ctr}(D')$.
%We fix a point $\mathsf{ctr}(A) \in A$ for each object $A \in S$, and call it the \textit{center} of $A$.
%We pick an arbitrary point $p(A,A') \in A \cap A'$ and then define $\sigma(A,A')$ as the polygonal curve obtained by concatenating $\sigma$ and $\sigma'$, where $\sigma$ (resp., $\sigma'$) is the segment connecting $p(A,A')$ and $\mathsf{ctr}(A)$ (resp., $\mathsf{ctr}(A')$).
%By construction, $\sigma(A,A')$ is a 2-piece polygonal curve connecting $\mathsf{ctr}(A)$ and $\mathsf{ctr}(A')$, which is contained in $A \cup A'$ because of the convexity of $A$ and $A'$.
%Now the centers $\mathsf{ctr}(A)$ of the objects together with the curves $\sigma(A,A')$ form a drawing of $G$ in the plane (which possibly has crossings).
For a path $\phi = (D_1,\dots,D_r)$ in $G[\mathcal{D}]$, we define $\gamma_\phi$ as the curve obtained by concatenating the curves $\sigma(D_1,D_2),\dots,\sigma(D_{r-1},D_r)$ in order.
If $\phi$ is a cycle, i.e., $D_1 = D_r$, then $\gamma_\phi$ is a closed curve.
The following lemma gives a characterization of the subsets of $\mathcal{D}$ enclosing a point.

\begin{lemma} \label{lem-characterization}
Let $\mathcal{D}_0 \subseteq \mathcal{D}$ and $x \in \mathbb{R}^2$.
The following statements are equivalent.
\begin{enumerate}[(i)]
    \item $\mathcal{D}_0$ encloses $x$.
    \item $G[\mathcal{D}_0]$ contains a cycle $\phi$ such that $\gamma_\phi$ is non-contractible in $\mathbb{R}^2 \backslash \{x\}$.
    \item $\{\sigma(D,D'): (D,D') \in \mathsf{Int}(\mathcal{D}_0)\}$ encloses $x$.
\end{enumerate}
%A subset $S' \subseteq S$ encloses a point $x \in \mathbb{R}^2$ iff $G[S']$ contains a cycle $\phi$ such that the closed curve $\gamma_\phi$ is non-contractible in $\mathbb{R}^2 \backslash \{x\}$.
\end{lemma}
\begin{proof}
Clearly, (iii) implies (i) since $\bigcup_{(D,D') \in \mathsf{Int}(\mathcal{D}_0)} \sigma(D,D') \subseteq \bigcup_{D \in \mathcal{D}_0} D$.
We shall show that (i) $\Rightarrow$ (ii) $\Rightarrow$ (iii).
To see (i) implies (ii), suppose $\mathcal{D}_0$ encloses $x$.
Let $U$ denote the union of all unit disks in $\mathcal{D}_0$.
The outer boundary of $U$ consists of several simple curves; one of these curves encloses $x$ and we denote it by $\xi$.
Note that $\xi$ is non-contractible in $\mathbb{R}^2 \backslash \{x\}$ by Jordan curve theorem.
Suppose $\xi$ consists of circular arcs $\xi_1,\dots,\xi_r$ (sorted in the order they appear on $\xi$).
Set $\xi_0 = \xi_r$ for convenience.
Let $D_0,D_1,\dots,D_r$ be the unit disks in $\mathcal{D}_0$ contributing the circular arcs $\xi_0,\xi_1,\dots,\xi_r$, respectively.
Then $\phi = (D_0,D_1,\dots,D_r)$ is a cycle in $G[\mathcal{D}_0]$.
It suffices to show that $\gamma_\phi$ is non-contractible in $\mathbb{R}^2 \backslash \{x\}$.
%$\gamma_\phi$ is homotopic to $\xi$ in $\mathbb{R}^2 \backslash \{x\}$, and is thus non-contractible in $\mathbb{R}^2 \backslash \{x\}$.
For $i \in [r]$, let $p_i$ be the intersection point of $\xi_{i-1}$ and $\xi_i$, and $\mu_i$ (resp., $\mu_i'$) be the curve from $p_i$ to $\mathsf{ctr}(D_i)$ (resp., from $\mathsf{ctr}(D_i)$ to $p_i$) defined by the segment between $\mathsf{ctr}(D_i)$ and $p_i$.
Consider the closed curve $\eta$ obtained by concatenating $\xi_0,\mu_1,\mu_1',\xi_1,\mu_2,\mu_2',\xi_2\dots,\xi_{r-1},\mu_r,\mu_r'$ in order.
Clearly, $\eta$ is homotopic to $\xi$ in $\mathbb{R}^2 \backslash \{x\}$ (when picking the same point as the base points of $\eta$ and $\xi$), since concatenating $\mu_i$ and $\mu_i'$ results in a contractible closed curve for every $i \in [r]$.
Since $\xi$ is non-contractible in $\mathbb{R}^2 \backslash \{x\}$, so is $\eta$.
On the other hand, we observe that $\eta$ is homotopic to $\gamma_\phi$ in $\mathbb{R}^2 \backslash \{x\}$.
To see this, let $\eta_i$ be the curve from $\mathsf{ctr}(D_i)$ to $\mathsf{ctr}(D_{i+1})$ obtained by concatenating $\mu_i',\xi_i,\mu_{i+1}$ in order, for $i \in [r]$.
Then $\eta$ is the concatenation of $\eta_1,\dots,\eta_r$.
Now note that $\eta_i$ is homotopic to $\sigma(D_i,D_{i+1})$ in $D_i \cup D_{i+1}$ (and hence in $\mathbb{R}^2 \backslash \{x\}$), as $D_i \cup D_{i+1}$ is simply-connected.
Thus, $\eta$ is homotopic to $\gamma_\phi$ in $\mathbb{R}^2 \backslash \{x\}$, which implies that $\gamma_\phi$ is also non-contractible in $\mathbb{R}^2 \backslash \{x\}$.

To see (ii) implies (iii), suppose $G[\mathcal{D}_0]$ contains a cycle $\phi$ such that $\gamma_\phi$ is non-contractible in $\mathbb{R}^2 \backslash \{x\}$.
It suffices to show that the image of $\gamma_\phi$ encloses $x$, since the image of $\gamma_\phi$ is contained in $\bigcup_{(D,D') \in \mathsf{Int}(\mathcal{D}_0)} \sigma(D,D')$.
Consider a point $o \in \mathbb{S}^2$ on the sphere $\mathbb{S}^2$.
Let $f: \mathbb{R}^2 \rightarrow \mathbb{S}^2 \backslash \{o\}$ be the natural homeomorphism.
Here $o$ is viewed as the point at infinity of the plane.
The curve $\gamma_\phi$ corresponds to a curve on $\mathbb{S}^2 \backslash \{o\}$ by the homeomorphism $f$, which we denote by $\gamma_\phi'$.
Assume $x$ is not enclosed by the image of $\gamma_\phi$.
Then on $\mathbb{S}^2$ there exists a simple curve $\tau$ connecting $f(x)$ and $o$, which is disjoint from the image of $\gamma_\phi'$.
Note that $\mathbb{S}^2 \backslash \tau$ is simply-connected.
Therefore, $\gamma_\phi'$ is contractible in $\mathbb{S}^2 \backslash \tau$ and hence contractible in $\mathbb{S}^2 \backslash \{x,o\}$.
It follows that $\gamma_\phi$ is contractible in $\mathbb{R}^2 \backslash \{x\}$, contradicting our assumption.
\end{proof}

\subsection{Sparsification step}

We construct a grid $\varGamma$ in the plane with $\frac{1}{2} \times \frac{1}{2}$ square cells.
For each cell $\Box \in \varGamma$, we write $\mathcal{D}_\Box = \{D \in \mathcal{D}: \mathsf{ctr}(D) \in \Box\}$.
For $\mathcal{D}' \subseteq \mathcal{D}$, let $\mathsf{opt}(X,\mathcal{D}')$ denote the minimum number of obstacles in $\mathcal{D}'$ needed to enclose $X$.
The goal of this section is to prove the following lemma.
%where $\mathsf{opt}(X,\mathcal{D}')$ (resp., $\mathsf{opt}(X,\mathcal{D})$) denote the minimum number of obstacles in $\mathcal{D}'$ (resp., $\mathcal{D}$) needed to enclose $X$.

\begin{lemma} \label{lem-sparse}
There exists a constant $c > 0$ such that one can compute in polynomial time a subset $\mathcal{D}' \subseteq \mathcal{D}$ satisfying $|\mathcal{D}' \cap \mathcal{D}_\Box| \leq c$ for all $\Box \in \varGamma$ and $\mathsf{opt}(X,\mathcal{D}') \leq c \cdot \mathsf{opt}(X,\mathcal{D})$.
\end{lemma}

A pair $(\Box,\Box')$ of grid cells in $\varGamma$ is \textit{relevant} if there exist $D \in \mathcal{D}_\Box$ and $D' \in \mathcal{D}_{\Box'}$ such that $D \cap D' \neq \emptyset$.
Note that for each $\Box \in \varGamma$, there are $O(1)$ cells $\Box' \in \varGamma$ such that $(\Box,\Box')$ is relevant.
To construct the desired $\mathcal{D}'$ in Lemma~\ref{lem-sparse}, we consider all relevant pairs.
For each relevant pair $(\Box,\Box')$, we include in $\mathcal{D}'$ some unit disks in $\mathcal{D}_\Box \cup \mathcal{D}_{\Box'}$ as follows.
Let $Y \subseteq X$ be the set of points enclosed by $\mathcal{D}_\Box \cup \mathcal{D}_{\Box'}$.
Also, let $\mathsf{Int}^* \subseteq \mathsf{Int}(\mathcal{D}_\Box \cup \mathcal{D}_{\Box'})$ consist of the edges $(D,D')$ where $D \in \mathcal{D}_\Box$ and $D' \in \mathcal{D}_{\Box'}$.
If $Y = \emptyset$, we arbitrarily pick $(D,D') \in \mathsf{Int}^*$, and include $D,D'$ in $\mathcal{D}'$.
If $Y \neq \emptyset$, we shall show the existence of at most four unit disks in $\mathcal{D}_\Box \cup \mathcal{D}_{\Box'}$ which enclose all points in $Y$; then we include them in $\mathcal{D}'$.
%\begin{observation}
%If $Y \neq \emptyset$, then $\mathsf{dist}(\Box,\Box') \geq 2 - \sqrt{2}$.
%\end{observation}

\begin{figure}[h]
    \centering
    \includegraphics[height=3.5cm]{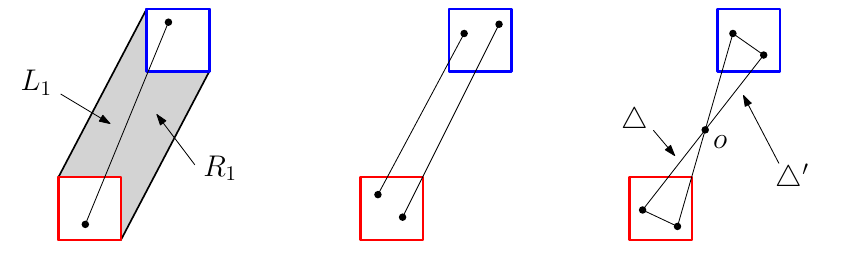}
    \caption{Illustration of the proof of Observation~\ref{obs-twopairs} --- $\Box$ is red, $\Box'$ is blue, and $Q \backslash (\Box \cup \Box')$ is grey.}
    \label{fig-twobox}
\end{figure}

%An \textit{intersecting pair} refers to a pair $(D,D')$ of intersecting unit disks where $D \in \mathcal{D} \cap \Box$ and $D' \in \mathcal{D} \cap \Box'$.
%Consider two unit disks $D \in \mathcal{D} \cap \Box$ and $D' \in \mathcal{D} \cap \Box'$ satisfying $D \cap D' \neq \emptyset$.
Consider an edge $(D,D') \in \mathsf{Int}^*$.
The line containing the segment $\sigma(D,D')$ partitions the plane into two halfplanes $H_L$ and $H_R$, where $H_L$ (resp., $H_R$) is to the left (resp., right) of the line with respect to the direction from $\mathsf{ctr}(D)$ to $\mathsf{ctr}(D')$.
Define $Y_L(D,D') = Y \cap H_L$ and $Y_R(D,D') = Y \cap H_R$.
We have the following key observation.
%Then we define a partial order $\leq_L$ among all intersecting pairs where $(D,D') \leq_L (E,E')$ if $Y_L(D,D') \subseteq Y_L(E,E')$.
%Similarly, we can also define another partial order $\leq_R$ where $(D,D') \leq_R (E,E')$ if $Y_R(D,D') \subseteq Y_R(E,E')$.
\begin{observation} \label{obs-twopairs}
Let $(D,D'),(E,E') \in \mathsf{Int}^*$.
\begin{enumerate}[(i)]
    \item Either $Y_L(D,D') \subseteq Y_L(E,E')$ or $Y_L(E,E') \subseteq Y_L(D,D')$.
    \item Either $Y_R(D,D') \subseteq Y_R(E,E')$ or $Y_R(E,E') \subseteq Y_R(D,D')$.
    %\item For any point $p \in Y$, $p$ is enclosed by $\{D,D',E,E'\}$ iff $p \in Y_L(D,D') \cap Y_R(E,E')$ or $p \in Y_L(E,E') \cap Y_R(D,D')$.
\end{enumerate}
\end{observation}
\begin{proof}
%It suffices to prove that either $Y_L(D,D') \subseteq Y_L(E,E')$ or $Y_L(E,E') \subseteq Y_L(D,D')$.
Let $Q$ be the convex hull of $\Box \cup \Box'$.
We first show that $Y \subseteq Q \backslash (\Box \cup \Box')$.
Note that $Y \cap (\Box \cup \Box') = \emptyset$, since $\Box$ is contained in any unit disk in $\mathcal{D}_\Box$ and $\Box'$ is contained in any unit disk in $\mathcal{D}_\Box'$.
So it suffices to show $Y \subseteq Q$.
Suppose there exists a point $a \in Y \backslash Q$.
Since $Q$ is convex, there exists a line $\ell$ such that $a$ and $Q$ lie on different sides of $\ell$.
Consider the ray $\psi$ shot from $a$ that is perpendicular to $\ell$ and does not intersect $\ell$.
For any point $b$ lying on the other side of $\ell$ than $a$, we have $\mathsf{dist}(a,b) \geq \mathsf{dist}(a',b)$ for all $a' \in \psi$.
Therefore, if a unit disk centered at the other side of $\ell$ than $a$ does not contain $a$, then it does not contain any point on $\psi$ and is thus disjoint from $\psi$.
Since $a$ is not contained in any unit disk in $\mathcal{D}_\Box \cup \mathcal{D}_{\Box'}$ and the centers of the unit disks in $\mathcal{D}_\Box \cup \mathcal{D}_{\Box'}$ all lie on the other side of $\ell$ than $a$, we know that $\psi$ is disjoint from all unit disks in $\mathcal{D}_\Box \cup \mathcal{D}_{\Box'}$.
This implies that $a$ is not enclosed by $\mathcal{D}_\Box \cup \mathcal{D}_{\Box'}$, which contradicts the fact $a \in Y$.
Thus, $Y \subseteq Q$ and $Y \subseteq Q \backslash (\Box \cup \Box')$.
%Furthermore, $Y \cap (\Box \cup \Box') = \emptyset$, because $\Box$ is contained in any unit disk in $\mathcal{D} \cap \Box$ and $\Box'$ is contained in any unit disk in $\mathcal{D} \cap \Box'$.
%It follows that $Y \subseteq Q \backslash (\Box \cup \Box')$.

To prove the lemma, notice that (i) and (ii) are symmetric.
Thus, it suffices to show (i).
The segment $\sigma(D,D')$ partitions the region $Q \backslash (\Box \cup \Box')$ into two parts, the left part $L_1$ and the right part $R_1$ (with respect to the direction from $\mathsf{ctr}(D)$ to $\mathsf{ctr}(D')$), where $L_1$ contains $Y_L(D,D')$.
See the left figure of Figure~\ref{fig-twobox}.
Similarly, $\sigma(E,E')$ also partitions $Q \backslash (\Box \cup \Box')$ into the left part $L_2$ and the right part $R_2$, where $L_2$ contains $Y_L(E,E')$.
%We distinguish two cases, depending on whether $\sigma(D,D') \backslash (\Box \cup \Box')$ and $\sigma(E,E') \backslash (\Box \cup \Box')$ intersect or not.
If $\sigma(D,D') \backslash (\Box \cup \Box')$ and $\sigma(E,E') \backslash (\Box \cup \Box')$ do not intersect, then either $L_1 \subseteq L_2$ and or $L_2 \subseteq L_1$.
See the middle figure of Figure~\ref{fig-twobox}.
%Without loss of generality, assume $L_1 \subseteq L_2$.
We then have either $Y_L(D,D') \subseteq Y_L(E,E')$ or $Y_L(E,E') \subseteq Y_L(D,D')$, which implies (i).
Next, suppose $\sigma(D,D') \backslash (\Box \cup \Box')$ and $\sigma(E,E') \backslash (\Box \cup \Box')$ intersect at the point $o$.
Let $\triangle$ be the triangle with vertices $\mathsf{ctr}(D),\mathsf{ctr}(E),o$ and $\triangle'$ be the triangle with vertices $\mathsf{ctr}(D'),\mathsf{ctr}(E'),o$.
See the right figure of Figure~\ref{fig-twobox}.
One of $\triangle$ and $\triangle'$ contains $L_1 \backslash L_2$, while the other one contains $L_2 \backslash L_1$.
Since $o \in \sigma(D,D')$, we have $o \in D$ or $o \in D'$; without loss of generality, assume $o \in D$.
Then we have $\triangle \subseteq D$, because all the three vertices of $\triangle$ are contained in $D$.
It follows that $Y \cap \triangle = \emptyset$, since no point in $X$ lies in $D$.
Therefore, we have either $Y \cap (L_1 \backslash L_2) = \emptyset$ (if $\triangle$ contains $L_1 \backslash L_2$) or $Y \cap (L_2 \backslash L_1) = \emptyset$ (if $\triangle$ contains $L_2 \backslash L_1$).
The former implies $Y_L(D,D') \subseteq Y_L(E,E')$ while the latter implies $Y_L(E,E') \subseteq Y_L(D,D')$.
\end{proof}

By (i) of the above observation, there exists $(D_L,D_L') \in \mathsf{Int}^*$ such that $Y_L(D,D') \subseteq Y_L(D_L,D_L')$ for $(D,D') \in \mathsf{Int}^*$.
Similarly, by (ii) of the above observation, there exists $(D_R,D_R') \in \mathsf{Int}^*$ such that $Y_R(D,D') \subseteq Y_R(D_R,D_R')$ for all $(D,D') \in \mathsf{Int}^*$.
We then include the four unit disks $D_L,D_L',D_R,D_R'$ in $\mathcal{D}'$.
\begin{observation} \label{obs-4disks}
    All points in $Y$ are enclosed by $\{D_L,D_L',D_R,D_R'\}$.
\end{observation}
\begin{proof}
Consider a point $y \in Y$.
We first show that $y \in Y_L(D_L,D_L')$ and $y \in Y_R(D_R,D_R')$.
Without loss of generality, we only need to show $y \in Y_L(D_L,D_L')$.
By the choice of $(D_L,D_L')$, we have $Y_L(D,D') \subseteq Y_L(D_L,D_L')$ for all $(D,D') \in \mathsf{Int}^*$.
Therefore, it suffices to show the existence of $(D,D') \in \mathsf{Int}^*$ such that $y \in Y_L(D,D')$.
Since $y$ is enclosed by $\mathcal{D}_\Box \cup \mathcal{D}_{\Box'}$, it is also enclosed by $\{\sigma(D,D'): (D,D') \in \mathsf{Int}(\mathcal{D}_\Box \cup \mathcal{D}_{\Box'})\}$ by Lemma~\ref{lem-characterization}.
As observed in the proof of Observation~\ref{obs-twopairs}, we have $y \in Q \backslash (\Box \cup \Box')$ where $Q$ is the convex hull of $\Box \cup \Box'$.
The segments $\sigma(D,D')$ for $(D,D') \in \mathsf{Int}^*$ decompose the region $Q \backslash (\Box \cup \Box')$ into small faces, among which there are two faces incident to the boundary of $Q$ (which we call \textit{boundary faces}).
If $y \in Y_R(D,D')$ for all $(D,D') \in \mathsf{Int}^*$, then $y$ lies in a boundary face.
In this case, $y$ is not enclosed by $\{\sigma(D,D'): (D,D') \in \mathsf{Int}^*\} \cup \{\Box,\Box'\}$, and thus not enclosed by $\{\sigma(D,D'): (D,D') \in \mathsf{Int}(\mathcal{D}_\Box \cup \mathcal{D}_{\Box'})\}$ because $\sigma(D,D') \subseteq \Box \cup \Box'$ for all $(D,D') \in \mathsf{Int}(\mathcal{D}_\Box \cup \mathcal{D}_{\Box'}) \backslash \mathsf{Int}^*$.
But this contradicts the fact that $y \in Y$.
Therefore, $y \in Y_L(D,D')$ for some $(D,D') \in \mathsf{Int}^*$ and hence $y \in Y_L(D_L,D_L')$.

To further see that $y$ is enclosed by $\{D_L,D_L',D_R,D_R'\}$, observe that $y$ is enclosed by 
\begin{equation*}
    \{\sigma(D_L,D_L'),\sigma(D_R,D_R'),\Box,\Box'\},    
\end{equation*}
since $y \in Q \backslash (\Box \cup \Box')$ and $y \in Y_L(D_L,D_L') \cap Y_R(D_R,D_R')$.
We have $\sigma(D_L,D_L') \subseteq D_L \cup D_L'$, $\sigma(D_R,D_R') \subseteq D_R \cup D_R'$, $\Box \subseteq D_L$, and $\Box' \subseteq D_L'$.
Hence, $y$ is enclosed by $\{D_L,D_L',D_R,D_R'\}$.
\end{proof}

Our construction of $\mathcal{D}'$ satisfies $|\mathcal{D}' \cap \mathcal{D}_\Box| = O(1)$ for all $\Box \in \varGamma$, simply because every grid cell in $\varGamma$ is only involved in a constant number of relevant pairs.
The following observation further shows that $\mathsf{opt}(X,\mathcal{D}') = O(\mathsf{opt}(X,\mathcal{D}))$, which completes the proof of Lemma~\ref{lem-sparse}.

\begin{observation}
    $\mathsf{opt}(X,\mathcal{D}') \leq c \cdot \mathsf{opt}(X,\mathcal{D})$, where $c = \max_{\Box \in \varGamma} |\mathcal{D}' \cap \mathcal{D}_\Box|$.
\end{observation}
\begin{proof}
Consider a minimum subset $\mathcal{D}_\mathsf{opt} \subseteq \mathcal{D}$ enclosing $X$.
We construct a subset $\mathcal{D}_\mathsf{opt}' \subseteq \mathcal{D}'$ as follows.
For every $\Box \in \varGamma$ with $\mathcal{D}_\mathsf{opt} \cap \mathcal{D}_\Box \neq \emptyset$, we include in $\mathcal{D}_\mathsf{opt}'$ all unit disks in $\mathcal{D}' \cap \mathcal{D}_\Box$.
Clearly, we have $|\mathcal{D}_\mathsf{opt}'| \leq c \cdot |\mathcal{D}_\mathsf{opt}|$.
It suffices to show that $\mathcal{D}_\mathsf{opt}'$ encloses $X$.

Consider a point $x \in X$.
We say a cycle $\phi = (D_0,D_1,\dots,D_r)$ in $G[\mathcal{D}_\mathsf{opt} \cup \mathcal{D}_\mathsf{opt}']$ is \textit{good} if it satisfies (i) $\gamma_\phi$ is non-contractible in $\mathbb{R}^2 \backslash \{x\}$ and (ii) $r$ is even and for every $i \in [\frac{r}{2}]$ there exists $\Box_i \in \varGamma$ such that $D_{2i-1}, D_{2i} \in \mathcal{D}_{\Box_i}$.
The \textit{cost} of $\phi$ is defined as the number of indices $i \in [r]$ satisfying $D_i \notin \mathcal{D}_\mathsf{opt}'$.
We claim that $G[\mathcal{D}_\mathsf{opt} \cup \mathcal{D}_\mathsf{opt}']$ contains at least one good cycle.
Since $\mathcal{D}_\mathsf{opt}$ encloses $x$, by Lemma~\ref{lem-characterization}, there exists a cycle $\phi = (D_0,D_1,\dots,D_r)$ in $G[\mathcal{D}_\mathsf{opt}]$ (and thus in $G[\mathcal{D}_\mathsf{opt} \cup \mathcal{D}_\mathsf{opt}']$) such that $\gamma_\phi$ is non-contractible in $\mathbb{R}^2 \backslash \{x\}$.
This cycle satisfies condition (i) for good cycles but not condition (ii).
Now define $\psi = (D_0,D_0,D_1,D_1,\dots,D_r,D_r)$, which is a cycle in $G[\mathcal{D}_\mathsf{opt} \cup \mathcal{D}_\mathsf{opt}']$ satisfying condition (ii).
Note that $\psi$ also satisfies condition (i) since $\gamma_\psi = \gamma_\phi$.
Thus, $\psi$ is good and $G[\mathcal{D}_\mathsf{opt} \cup \mathcal{D}_\mathsf{opt}']$ contains at least one good cycle.

To prove that $\mathcal{D}_\mathsf{opt}'$ encloses $x$, let $\phi = (D_0,D_1,\dots,D_r)$ be a good cycle in $G[\mathcal{D}_\mathsf{opt} \cup \mathcal{D}_\mathsf{opt}']$ with minimum cost.
By condition (i), $\gamma_\phi$ is non-contractible in $\mathbb{R}^2 \backslash \{x\}$.
If $D_1,\dots,D_r \in \mathcal{D}_\mathsf{opt}'$, then we are done.
Indeed, in this case, $\phi$ is also a cycle in $G[\mathcal{D}_\mathsf{opt}']$ and Lemma~\ref{lem-characterization} implies that $\mathcal{D}_\mathsf{opt}'$ encloses $x$.
So suppose $D_j \notin \mathcal{D}_\mathsf{opt}'$ for some $j \in [r]$.
We assume $j = 2i-1$ for some $i \in [\frac{r}{2}]$; the case where $j$ is even can be handled in the same way.
Since $\phi$ is good, by condition (ii), there exists $\Box_i \in \varGamma$ (resp., $\Box_{i-1} \in \varGamma$) such that $D_j,D_{j+1} \in \mathcal{D}_{\Box_i}$ (resp., $D_{j-2},D_{j-1} \in \mathcal{D}_{\Box_{i-1}}$).
Note that $(\Box_{i-1},\Box_i)$ is a relevant pair, because $D_{j-1} \cap D_j \neq \emptyset$.
Recall that when considering $(\Box_{i-1},\Box_i)$, we included in $\mathcal{D}'$ (at most) four unit disks in $\mathcal{D}_{\Box_{i-1}} \cup \mathcal{D}_{\Box_i}$, and among them there exist two intersecting unit disks $D_{j-1}' \in \mathcal{D}_{\Box_{i-1}}$ and $D_j' \in \mathcal{D}_{\Box_i}$.
We have $D_{j-1}',D_j' \in \mathcal{D}_\mathsf{opt}'$, since $\mathcal{D}_\mathsf{opt} \cap \mathcal{D}_{\Box_{i-1}} \neq \emptyset$ and $\mathcal{D}_\mathsf{opt} \cap \mathcal{D}_{\Box_i} \neq \emptyset$.
As $D_{j-2},D_{j-1}' \in \mathcal{D}_{\Box_{i-1}}$ and $D_j',D_{j+1} \in \mathcal{D}_{\Box_i}$, we have $D_{j-2} \cap D_{j-1}' \neq \emptyset$ and $D_j' \cap D_{j+1} \neq \emptyset$.
Therefore, $\phi' = (D_0,D_1,\dots,D_{j-2},D_{j-1}',D_j',D_{j+1},\dots,D_r)$ is also a cycle in $G[\mathcal{D}_\mathsf{opt} \cup \mathcal{D}_\mathsf{opt}']$.
Observe that $\phi'$ cannot be a good cycle.
Indeed, if $\phi'$ is good, the cost of $\phi'$ is strictly smaller than the cost of $\phi$ as $D_{j-1}',D_j' \in \mathcal{D}_\mathsf{opt}'$ but $D_j \notin \mathcal{D}_\mathsf{opt}'$, which contradicts the fact that $\phi$ is a good cycle with minimum cost.
However, $\phi'$ satisfies condition (ii) for good cycles, because $D_{j-2},D_{j-1}' \in \mathcal{D} \cap \mathcal{D}_{\Box_{i-1}}$ and $D_j',D_{j+1} \in \mathcal{D}_{\Box_i}$.
Thus, $\phi'$ does not satisfy condition (i), i.e., $\gamma_{\phi'}$ is contractible in $\mathbb{R}^2 \backslash \{x\}$.
It follows that $\gamma_\phi$ and $\gamma_{\phi'}$ are not homotopic in $\mathbb{R}^2 \backslash \{x\}$.
Now consider the paths $\psi = (D_{j-2},D_{j-1},D_j,D_{j+1})$ and $\psi' = (D_{j-2},D_{j-1}',D_j',D_{j+1})$ in $G[\mathcal{D}_\mathsf{opt} \cup \mathcal{D}_\mathsf{opt}']$.
Their corresponding curves $\gamma_\psi$ and $\gamma_{\psi'}$ share the same endpoints.
Note that if $\gamma_\psi$ and $\gamma_{\psi'}$ are homotopic in $\mathbb{R}^2 \backslash \{x\}$, then $\gamma_\phi$ and $\gamma_{\phi'}$ are also homotopic in $\mathbb{R}^2 \backslash \{x\}$ (when picking the same point as their base points).
Hence, $\gamma_\psi$ and $\gamma_{\psi'}$ are not homotopic in $\mathbb{R}^2 \backslash \{x\}$.
This further implies $\gamma_\xi$ is non-contractible in $\mathbb{R}^2 \backslash \{x\}$, where $\xi = (D_{j-2},D_{j-1},D_j,D_{j+1},D_j',D_{j-1}',D_{j-2})$ is a cycle in $G[\mathcal{D}_\mathsf{opt} \cup \mathcal{D}_\mathsf{opt}']$.
Since all vertices of $\xi$ are in $\mathcal{D}_{\Box_{i-1}} \cup \mathcal{D}_{\Box_i}$, by Lemma~\ref{lem-characterization}, $\mathcal{D}_{\Box_{i-1}} \cup \mathcal{D}_{\Box_i}$ encloses $x$.
Observation~\ref{obs-4disks} then implies that the four unit disks we include in $\mathcal{D}'$ for the relevant pair $(\Box_{i-1},\Box_i)$ also enclose $x$.
These unit disks are all in $\mathcal{D}_\mathsf{opt}'$ as $\mathcal{D}_\mathsf{opt} \cap \mathcal{D}_{\Box_{i-1}} \neq \emptyset$ and $\mathcal{D}_\mathsf{opt} \cap \mathcal{D}_{\Box_i} \neq \emptyset$.
Thus, $\mathcal{D}_\mathsf{opt}'$ encloses $x$.
%Without loss of generality, we can assume $r$ is even and for every $i \in [\frac{r}{2}]$ there exists a cell $\Box_i \in \varGamma$ such that $D_{2i-1}, D_{2i} \in \mathcal{D} \cap \Box$.
\end{proof}

\subsection{Approximation for the sparse case via min-cut}

Thanks to Lemma~\ref{lem-sparse}, we can now assume $|\mathcal{D}_\Box| = O(1)$ for all $\Box \in \varGamma$, and design a constant-approximation algorithm under this assumption.

First, we observe that the maximum degree of the graph $G[\mathcal{D}]$ is $O(1)$.
Let $D \in \mathcal{D}_\Box$.
For every neighbor $D'$ of $D$ in $G[\mathcal{D}]$, we must have $D' \in \mathcal{D}_{\Box'}$ for a cell $\Box' \in \varGamma$ with constant distance from $\Box$.
The number of such cells is $O(1)$ and $|\mathcal{D}_{\Box'}| = O(1)$ for each such cell $\Box'$.
Thus, $D$ has $O(1)$ neighbors and the maximum degree of $G[\mathcal{D}]$ is $O(1)$.

Consider the following drawing of the graph $G[\mathcal{D}]$ in the plane.
We draw each vertex $D \in \mathcal{D}$ at the point $\mathsf{ctr}(D)$.
Then we draw each edge $(D,D') \in \mathsf{Int}(\mathcal{D})$ as the segment $\sigma(D,D')$.
This drawing is not necessarily planar because (the images of) the edges can cross.
However, it has a nice and important property: each edge crosses with at most $O(1)$ other edges.
Again, this follows from the sparsity of $\mathcal{D}$.
Let $(D,D') \in \mathsf{Int}(\mathcal{D})$ where $D \in \mathcal{D}_\Box$ and $D' \in \mathcal{D}_{\Box'}$.
If $\sigma(D,D')$ crosses with $\sigma(E,E')$ for $(E,E') \in \mathsf{Int}(\mathcal{D})$, then the cells containing $\mathsf{ctr}(E)$ and $\mathsf{ctr}(E')$ must be with constant distance from $\Box$ and $\Box'$.
There are $O(1)$ such cells each of which contains $O(1)$ centers of the unit disks in $\mathcal{D}$.
Thus, the number of edges whose images cross with $\sigma(D,D')$ is $O(1)$.

\begin{figure}[h]
    \centering
    \includegraphics[height=3.5cm]{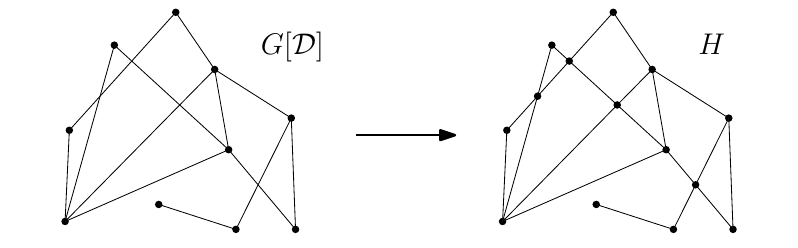}
    \caption{Illustration of the construction of $H$ from the drawing of $G[\mathcal{D}]$.}
    \label{fig-planarize}
\end{figure}

From the drawing of $G[\mathcal{D}]$, we create a planar graph $H$ as follows.
Let $C$ be the set of crossing points of the drawing of $G[\mathcal{D}]$.
The vertex set of $H$ is $\{\mathsf{ctr}(D): D \in \mathcal{D}\} \cup C$.
The points in $C$ subdivide the image of each edge of $G[\mathcal{D}]$ into $O(1)$ pieces.
These pieces, which have endpoints in $\{\mathsf{ctr}(D): D \in \mathcal{D}\} \cup C$, are the edges of $H$.
The drawing of $G[\mathcal{D}]$ induces a planar drawing of $H$.
See Figure~\ref{fig-planarize}.
%We add a new vertex at each crossing point of the drawing of $G[\mathcal{D}]$.
%These new vertices subdivide the image of each edge of $G[\mathcal{D}]$ into $O(1)$ pieces.
We have the following important observation.

\begin{observation}
There exists a subset $E \subseteq E(H)$ with $|E| = O(\mathsf{opt}(X,\mathcal{D}))$ which encloses $X$.
Furthermore, given a subset $E \subseteq E(H)$ that encloses $X$, one can compute in polynomial time a subset $\mathcal{D}_0 \subseteq \mathcal{D}$ enclosing $X$ such that $|\mathcal{D}_0| \leq 2|E|$.
\end{observation}
\begin{proof}
Let $\mathcal{D}_\mathsf{opt} \subseteq \mathcal{D}$ be a subset of size $\mathsf{opt}(X,\mathcal{D})$ that encloses $X$.
Since the maximum degree of $G[\mathcal{D}]$ is $O(1)$, $|\mathsf{Int}(\mathcal{D}_\mathsf{opt})| = O(\mathsf{opt}(X,\mathcal{D}))$.
Each edge $(D,D') \in \mathsf{Int}(\mathcal{D}_\mathsf{opt})$ corresponds to $O(1)$ edges in $H$ (which form a subdivision of $\sigma(D,D')$).
Let $E \subseteq E(H)$ consist of all edges in $H$ correspond to the edges in $\mathsf{Int}(\mathcal{D}_\mathsf{opt})$.
As $|\mathsf{Int}(\mathcal{D}_\mathsf{opt})| = O(\mathsf{opt}(X,\mathcal{D}))$, $|E| = O(\mathsf{opt}(X,\mathcal{D}))$.
The union of (the images) of the edges in $E$ is exactly equal to the union of $\sigma(D,D')$ for $(D,D') \in \mathsf{Int}(\mathcal{D}_\mathsf{opt})$, while the latter encloses $X$ by Lemma~\ref{lem-characterization}.
Thus, $E$ encloses $X$.

Next, suppose we are given $E \subseteq E(H)$ that encloses $X$.
Each $e \in E$ is a piece of an edge $f(e) \in \mathsf{Int}(\mathcal{D})$.
We simply define $\mathcal{D}_0 \subseteq \mathcal{D}$ as the subset consisting of the endpoints of $f(e)$ for all $e \in E$.
Clearly, $|\mathcal{D}_0| \leq 2|E|$.
The union of $\sigma(D,D')$ for $(D,D') \in \mathsf{Int}(\mathcal{D}_0)$ contains the union of the edges in $E$, and hence encloses $X$.
By Lemma~\ref{lem-characterization}, $\mathcal{D}_0$ encloses $X$.
\end{proof}

It now suffices to compute a minimum-size $E_\mathsf{opt} \subseteq E(H)$ that encloses $X$.
The above observation then implies that $|E_\mathsf{opt}| = O(\mathsf{opt}(X,\mathcal{D}))$ and one can compute in polynomial time a subset $\mathcal{D}_0 \subseteq \mathcal{D}$ enclosing $X$ such that $|\mathcal{D}_0| \leq 2|E_\mathsf{opt}| = O(\mathsf{opt}(X,\mathcal{D}))$.
We show that computing $E_\mathsf{opt}$ can be reduced to the following \textit{minimum $S$-$T$ cut} problem.
%Recall that in the min-cut problem, the input is an edge-weighted graph with two terminal vertices, and the goal is to remove from the graph a set of edges with minimum total weight such that the two terminal vertices belong to different connected components in the resulting graph. 

\begin{tcolorbox}[colback=gray!5!white,colframe=gray!75!black]
{\bf Minimum $S$-$T$ Cut} \\
\textbf{Input:} A graph $G$ and two disjoint sets $S,T \subseteq V(G)$. \\
\textbf{Output:} A minimum subset $E \subseteq E(G)$ such that $s$ and $t$ lie in different connected components of $G - E$ for every $s \in S$ and every $t \in T$.
\end{tcolorbox}

Before discussing the reduction, we first observe that the above minimum $S$-$T$ cut problem is polynomial-time solvable.
Indeed, we can add to the input graph $G$ a source vertex $s$ connecting to all vertices in $S$ and a target vertex $t$ connecting to all vertices in $T$.
Then we give the original edges of $G$ weights $1$ and give the edges incident to $s$ (resp., $t$) weights $\infty$.
Let $G^+$ be the resulting edge-weighted graph.
Now a minimum $S$-$T$ cut in the original $G$ is equivalent to a minimum-weight $s$-$t$ cut in $G^+$.
The latter can be computed in polynomial time by the well-known duality between min-cut and max-flow.

To reduce our problem to minimum $S$-$T$ cut, we consider the dual graph $H^*$ of the planar graph $H$.
Each vertex of $H^*$ corresponds to a face $f$ of $H$, which is called the \textit{dual vertex} of $f$ and is denoted by $f^*$.
Each edge of $H^*$ corresponds to an edge $e$ of $H$, which is called the \textit{dual edge} of $e$ and is denoted by $e^*$; here $e^*$ connects the dual vertices of the two faces of $H$ incident to $e$.
Let $o$ be the outer face of $H$.
We say a face of $H$ is \textit{nonempty} if it contains at least one point in $X$.
We have the following observation.

\begin{observation}
A subset $E \subseteq E(H)$ encloses $X$ iff for every nonempty face $f$ of $H$, $f^*$ and $o^*$ lie in different connected components of $H^* - \{e^*: e \in E\}$.
\end{observation}
\begin{proof}
To see the ``if'' direction, assume $E$ does not enclose $X$.
%for every nonempty face $f$ of $H$, $f^*$ and $o^*$ lie in different connected components of $H^* - \{e^*: e \in E\}$.
Let $x \in X$ be a point not enclosed by $E$, and $f$ be the (nonempty) face of $H$ containing $x$.
As $E$ does not enclose $x$, there exists a curve $\gamma$ in the plane connecting $x$ and a point $y$ in the outer face $o$ of $H$ that does not intersect any edge in $E$.
Without loss of generality, we may assume that $\gamma$ does not intersect any vertex of $H$.
Suppose the faces of $H$ visited by $\gamma$ in order are $f_1,\dots,f_r$ (where $f_1 = f$ and $f_r = o$), and when $\gamma$ enters $f_{i+1}$ from $f_i$ it goes across the edge $e_i$.
Now $e_1,\dots,e_{r-1} \notin E$.
Thus, there is a path in $H^* - \{e^*: e \in E\}$ from $f^*$ to $o^*$, which consists of the edges $e_1,\dots,e_{r-1}$.
So $f^*$ and $o^*$ lie in the same connected component of $H^* - \{e^*: e \in E\}$.
To see the ``only if'' direction, assume that $E$ encloses $X$.
Consider a nonempty face $f$ of $H$.
To see $f^*$ and $o^*$ lie in different connected components of $H^* - \{e^*: e \in E\}$, let $\pi$ be a path from $f^*$ to $o^*$ in $H^*$.
Our goal is to show that at least one edge on $\pi$ is in $\{e^*: e \in E\}$.
Suppose the edges on $\pi$ are $e_1^*,\dots,e_r^*$.
Then one can connect $f$ and $o$ by a curve $\gamma$ in the plane that does not intersect any edge of $H$ except $e_1,\dots,e_r$.
Since $E$ encloses $X$ and $f$ contains at least one point in $X$, $\gamma$ must intersect at least one edge in $E$.
Thus, $e_i \in E$ for some $i \in [r]$ and $e_i^* \in \{e^*: e \in E\}$.
\end{proof}

By the above observation, computing a minimum-size $E_\mathsf{opt} \subseteq E(H)$ enclosing $X$ is equivalent to computing a minimum-size $E_\mathsf{opt}^* \subseteq E(H^*)$ such that $f^*$ and $o^*$ lie in different connected components of $H^* - E_\mathsf{opt}^*$ for any nonempty face $f$ of $H$.
Note that the latter is in turn equivalent to the minimum $S$-$T$ cut instance on $H^*$ with $S = \{o^*\}$ and $T = \{f^*: f \text{ is a nonempty face of }H\}$.
Thus, $E_\mathsf{opt}$ can be computed in polynomial time, and we finally obtain our algorithm for unit disks.

\begin{theorem}
There exists a polynomial-time $O(1)$-approximation algorithm for \EnclosingPoints with unit disks, where $n$ is the total number of points and unit disks.
\end{theorem}

%% file: LPframe.tex
%In our approximation algorithms, the polygons may have weights, and we define the \emph{weighted winding number} of $q$ with respect to $P$ as the weight of $P$ times the \emph{winding number} of $q$ with respect to $P$.

\section{Approach based on LP rounding}
%Here we present approximation algorithms for enclosing points with colored geometric objects. Our approach works for disks, arbitrary line segments, and more generally curves that pairwise intersect $O(1)$ times. In the following, we explain our approach in terms of line segments as an example.

%A common technique for designing approximation algorithms for set cover related problems is to use LP rounding. The idea is to first design and solve an LP relaxation (i.e., getting a fractional solution), and then round the LP solution to get an integral solution. Our algorithm follows this line of approach, which requires nontrivial geometric insights for designing and rounding the LP, as we will show.

In this section, we present our algorithmic framework based on LP rounding.
We present our algorithm for line segments.
It generalizes to curves pairwise intersecting a constant number of times, and also general disks; see Appendix~\ref{app-framework1}.

\begin{figure}[!htbp]
    \centering
    \includegraphics[width=.55\textwidth]{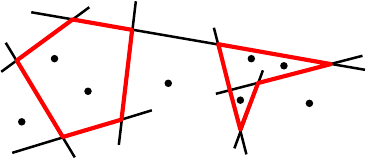}
    \caption{The outer boundary $\mathcal{B}$ of the union of segments is a set of disjoint simple polygons (shown in red).}
    \label{fig:outer_boundary_cycles}
\end{figure}

\paragraph{Formulation of the LP relaxation.}
Let $X$ be a set of points in $\mathbb{R}^2$ and $\mathcal{S}$ be a set of line segments.
We first formalize an LP relaxation of \EnclosingPoints.
Intuitively, the idea is as follows: a point $q$ is enclosed by a set $\mathcal{S}' \subseteq \mathcal{S}$ of segments iff the outer boundary of these segments enclose $q$.
So it suffices to specify the outer boundary $\mathcal{B}$ of the union of segments that we select in the solution.
$\mathcal{B}$ is a set of disjoint simple polygons (as shown in Fig.~\ref{fig:outer_boundary_cycles}), which can also be viewed as a set of cycles in a graph $G$: the vertices of $G$ are intersection points between two segments in $\mathcal{S}$, and the edges of $G$ are subsegments between a pair of vertices in $G$ that lie on the same segment in $\mathcal{S}$.
To get a fractional solution, our goal is to select a set of fractional cycles in $G$ to enclose all points.
%We can form the constraints for each color class $c$ separately.
% (TODO: after removing the segments on the corners)

We now formally describe our LP.
We start with specifying the variables in the LP.
%For each color class $c$, let $S_c$ denote the set of the input line segments with color $c$.
%The arrangement of $S_c$ can be viewed as a planar graph $Arr_c$, which has $O(|S_c|^2)$ vertices and edges. Each vertex in $Arr_c$ is an intersection point of two line segments in $S_c$.
The arrangement of $\mathcal{S}$ can be viewed as a planar graph $G_\mathsf{arr}$, which has $O(|\mathcal{S}|^2)$ vertices and edges.
Each vertex in $G_\mathsf{arr}$ is an intersection point of two line segments in $\mathcal{S}$.
We then create a new (directed) graph $G$, which has the same set of vertices as $G_\mathsf{arr}$.
For each line segment $\ell \in \mathcal{S}$, let $V(\ell)$ denote the set of all vertices in $G$ that lie on $\ell$.
For each pair of points $u,v\in V(\ell)$, we include in $G$ a directed edge $e_{uv}^{(\ell)}$ from $u$ to $v$, and also create a variable $x_{uv}^{(\ell)}$ where $0 \leq x_{uv}^{(\ell)} \leq 1$, indicating in what degree the directed subsegment $uv$ of $\ell$ (with direction from $u$ to $v$) is selected in the fractional solution.
We call $x_{uv}^{(\ell)}$ the weight of the directed subsegment $uv$.
%Also create a directed edge $e_{uv}^{(\ell)}$ from $u$ to $v$ in $G_c$.
%, and also including the two endpoints of $\ell$
%(In the special case, if $\ell$ does not intersect any other line segment, then we also create a variable $x_{00}^{(\ell)}$ to denote the whole line segment $\ell$.)

We emphasize that we need to create two variables (and respectively, two directed edges in $G$) for each undirected subsegment $uv$, because to ensure all points are enclosed according to the winding number constraints to be defined later, we need to know whether $uv$ appears on the boundary of the solution in clockwise or counter-clockwise order.

%\quad

We next present the constraints of our LP.
There are two types of constraints that need to be handled: flow constraints and winding number constraints.

\paragraph{Flow constraints.}
To ensure that the solution to the LP is a set of fractional cycles, we design the flow constraints as follows.
For each vertex $u$ of $G$, we create the constraint
\[\sum_{v,\ell:~e_{uv}^{(\ell)}\in E(G)} x_{uv}^{(\ell)}=\sum_{v,\ell:~e_{vu}^{(\ell)}\in E(G)} x_{vu}^{(\ell)}.\]
This guarantees that a solution to the LP is a circulation on $G$.
Using standard techniques, this circulation can be decomposed into $O(n^3)$ fractional cycles in polynomial time: repeatedly select the smallest variable $x_{uv}^{(\ell)}$ that is strictly positive, find a path $P_{vu}$ from $v$ to $u$ in $G$ using only edges with strictly positive values (such path always exists, and the value of each edge on the path is at least $x_{uv}^{(\ell)}$), and subtract from the solution the fractionally weighted cycle $C$ formed by concatenating $e_{uv}^{(\ell)}$ and $P_{vu}$.
The fractional cycle $C$ has weight $w_{\mathcal{C}}=x_{uv}^{(\ell)}$.
Each round sets the value of at least one variable $x_{uv}^{(\ell)}$ to $0$, so the process will terminate in $O(n^3)$ steps.
Each cycle in $G$ corresponds to a polygon (which is not necessarily simple) in the plane, so the LP solution can be decomposed into $O(n^3)$ fractionally weighted polygons.

\paragraph{Winding number constraints.}
To ensure that each point in $X$ is enclosed by the outer boundary of the union of the chosen subsegments, the idea is to constrain their winding number.
After orienting each (unknown) simple polygon on the outer boundary of the optimal solution $\mathcal{S}^*$ in counter-clockwise order, the winding number of each point $q \in X$ is exactly $1$.
So it suffices to ensure a similar inequality when we form the constraints for the fractional solution to the LP.

For each point $q\in X$, let $\vec{r}_q$ be an arbitrary ray starting from $q$.
Define $E_q^+ = \{(u,v): e_{uv}^{(\ell)}\text{ crosses }\vec{r}_q\text{ in counter-clockwise order}\}$ and $E_q^- = \{(u,v): e_{uv}^{(\ell)}\text{ crosses }\vec{r}_q\text{ in clockwise order}\}$.
Inspired by the winding-number algorithm, we add the following constraint to ensure the winding number of $q$ with respect to the LP solution that we selected is at least $1$:
\[\sum_{(u,v) \in E_q^+} x_{uv}^{(\ell)}-\sum_{(u,v) \in E_q^-} x_{uv}^{(\ell)}\geq 1,~\forall q\in X.\]
In other words, the (weighted) winding number of $q$ equals to the total weight of the edges that cross the ray $\vec{r}_q$ in counter-clockwise direction, minus the total weight of the edges that cross the ray $\vec{r}_q$ in clockwise direction.
We remark that we require the winding number of each point to be at least $1$ instead of exactly $1$, since in the LP solution, each polygon may not be simple, which means the winding numbers can be greater than $1$.

The winding number constraints together with the flow constraints ensure that each point is enclosed by the fractional cycles in the LP solution.
Namely, for each point $q\in X$, we have
\[\sum_{\text{cycle }C} w_C\cdot \wind(q,C)\geq 1.\]
%3. Constraint on $x_{uv}$. For each edge $(u,v)$, we form the constraint $0\leq x_{uv}\leq 1$.

The objective function to be minimized is just $\sum_{(u,v) \in E(G)} x_{uv}^{(\ell)}$, the total weight of the (fractionally) selected subsegments.

\paragraph{Rounding the LP.}
After solving the LP to compute the optimal fractional solution in polynomial time, we represent the LP solution as a set $\mathcal{C}$ of $O(n^2)$ fractionally weighted cycles in the graph $G$ as discussed above, and use these cycles to guide our rounding procedure.

%The rounding scheme is somehow standard, and is commonly used for approximating set cover solutions \cite{DBLP:books/daglib/0004338}.
%Here we provide a simple self-contained description, which is by using randomized rounding~\cite{RaghavanT87}. 
Our rounding scheme is similar to the randomized rounding used in~\cite{RaghavanT87} (which is commonly used for approximating set cover solutions \cite{DBLP:books/daglib/0004338}).
Basically speaking, for each fractional cycle $C \in \mathcal{C}$ of weight $w_C$, we randomly and independently select it with probability $\min\{10w_\mathcal{C}\log n,1\}$.
Here ``selecting a cycle'' means selecting all subsegments corresponding to the edges on the cycle, and if a subsegment $e_{uv}^{(\ell)}$ of an input line segment $\ell$ is selected, then the whole segment $\ell$ is included in our integral solution.
Note that a line segment $\ell$ may be included because of multiple subsegments of it are selected during the rounding process; although it suffices to include $\ell$ just once, in the analysis below, we treat as if $\ell$ would be included multiple times.
We show this will not affect the approximation factor by much, as the combinatorial complexity of the outer boundary of the objects we consider is near linear.

Our goal is to ensure all points in $X$ have winding number at least $1$ w.h.p. after the rounding process, which implies that they are enclosed by the integral solution.
But a technical issue arises for cycles in $\mathcal{C}$ that are not necessarily simple: they may wind around a point multiple times, resulting in a large positive winding number for the point.
In this case, even giving a small weight to the cycle can still ensure the (weighted) winding number of that point is at least $1$ in the fractional solution to the LP, but that means the cycle will only be selected with a small probability during the rounding process (selecting the cycle with a larger probability would be too costly).

To handle the above issue, the idea is to ``unwind'' the non-simple cycles in $\mathcal{}$ first before the rounding, i.e., decompose them into a set of simple cycles, and then randomly select each simple cycle independently.
This ensures that each cycle contributes at most once to the winding number of any point $q$.

\begin{figure}[!htbp]
    \centering
    \includegraphics[width=.4\textwidth]{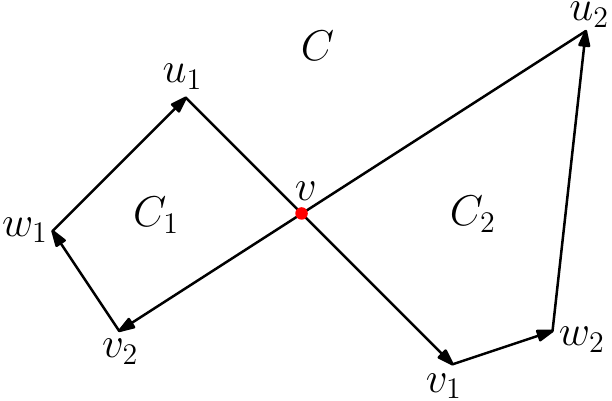}
    \caption{During the unwinding process, the non-simple cycle $C=u_1v_1w_2u_2v_2w_1$ with length 6 is decomposed into two simple cycles $C_1=vv_2w_1u_1$ and $C_2=vv_1w_2u_2$ each with length 4.}
    \label{fig:unwinding}
\end{figure}

To unwind a non-simple cycle $C \in \mathcal{C}$, we repeatedly choose a self-intersection point of $C$ and split $C$ into two cycles at that point.
%Suppose $v$ is the intersection point of a pair of crossing edges $u_1v_1$ and $u_2v_2$ of $C$.
Specifically, if two edges $u_1v_1$ and $u_2v_2$ of $C$ cross each other, then we create a new vertex $v$ at their intersection point, subdividing $u_1v_1$ (resp., $u_2v_2$) into two edges $u_1v$ and $vv_1$ (resp., $u_2v$ and $vv_2$).
In this way, we decompose the cycle $C$ into two cycles $C_1$ and $C_2$ by splitting at $v$ (see Figure~\ref{fig:unwinding}).
The total length of the cycles increases by $2$, since the two crossing edges are split into four edges. Recursively subdivide the cycles $C_1$ and $C_2$ if they are not simple.
%An example for the unwinding process is shown in Fig.~\ref{fig:unwinding}.
%The total running time is polynomial.
%the number of newly created arcs satisfy the recursion f(n)=f(n_1)+f(n-n_1+O(1)) for some n_1 (the length of the first cycle) => f(n)=O(n)

To upper bound the approximation ratio, we show that it only loses a constant approximation factor at the step of decomposing non-simple cycles into simple cycles.
Let $f(n)$ denote the maximum total number of edges that a non-simple cycle of length $n$ can decompose into, which satisfies the recurrence $f(n)\leq \max_{3\leq n_1\leq n-1} \left(f(n_1)+f(n-n_1+2)\right)$,
since a simple cycle formed by line segments would have length at least $3$.
The base case is $f(n)=n$ for $n\leq 3$.
Solving the recurrence, we get $f(n)\leq 3n-6=O(n)$.

Let $\mathcal{C}'$ be the set of fractional simple cycles obtained by unwinding the cycles in $\mathcal{C}$.
It still holds that for each point $q\in X$,
\[\sum_{\text{cycle }C \in \mathcal{C}'} w_C\cdot \wind(q,C) = \sum_{\text{cycle }C \in \mathcal{C}} w_C\cdot \wind(q,C) \geq 1.\]
We can ignore the simple cycles in $\mathcal{C}'$ that appear in clockwise order, because they only contribute negatively to the winding number constraints.
In other words, we can assume without loss of generality that all cycles in $\mathcal{C}'$ are counter-clockwise.

Now we analyze the probability of failure. 
For each $q \in X$, let $\mathcal{C}_q \subseteq \mathcal{C}'$ consist of the cycles that enclose $q$.
We then have
%After the rounding process, for any $q\in X$,
\begin{flalign*}
\Pr[q\text{ is not enclosed}] &=\prod_{C \in \mathcal{C}_q} \Pr[C\text{ is not selected}]\\
&=\prod_{C \in \mathcal{C}_q}\left(1-\min\{10w_C\log n,1\}\right) \\
&\leq \lim_{k\rightarrow \infty}\left(1-\frac{10\log n}{k}\right)^{k} \leq e^{-10\log n}\leq \frac{1}{n^{10}}.
\end{flalign*}
By union bound, the probability of existing any point $q\in X$ that is not enclosed is at most
\[\Pr[\exists q\in X,~q\text{ is not enclosed}]\leq n\cdot \frac{1}{n^{10}}\leq \frac{1}{n^{9}}.\]

Therefore, the rounding process succeeds w.h.p. We remark that the approach can be derandomized, by the standard method of conditional probabilities \cite{Raghavan88}.

\paragraph{Analyzing the approximation ratio.}
It is easy to see that the optimal solution $\mathcal{S}^*$ corresponds to a feasible solution to the LP: if we set the variables corresponding to the directed subsegments on the outer boundary of the arrangement of $\mathcal{S}^*$ to be $1$, and all other variables to be $0$, we will obtain a feasible solution to the LP. From the known combinatorial complexity of the outer face in the arrangement of line segments, the outer boundary of $\mathcal{S}^*$ can be decomposed into $O(|\mathcal{S}^*|\cdot \alpha(|\mathcal{S}^*|))$ subsegments.
Therefore, the optimal LP value satisfies $\opt = O(|\mathcal{S}^*|\cdot \alpha(|\mathcal{S}^*|))$.

As we have shown before, after the unwinding step, the total weight of edges among all cycles only increases by a constant factor, i.e., $\sum_{C \in \mathcal{C}'} w_C \cdot \mathsf{len}(C) = O(\opt)$, where $\mathsf{len}(C)$ is the length of $C$.
So after performing randomized rounding on the unwound fractional solution, the expected size of the resulting integral solution is at most $\sum_{C \in \mathcal{C}'} 10w_C\log n \cdot \mathsf{len}(C) = O(\opt \log n) = O(|\mathcal{S}^*|\cdot \alpha(|\mathcal{S}^*|) \log n)$.
So we conclude the following.

%In summary, the problem of enclosing points by geometric objects can be viewed as selecting the smallest subset of objects to cover all curves connecting a point $q$ with infinity (i.e.\ a set cover problem). The difficulty is there are infinitely many such curves, so directly applying the known set cover results does not give us a good approximation. Using the winding number idea, we transform the problem into covering each of a finite number of rays with a finite number of possible boundary curves of the input objects, so that we can apply the set cover approximation techniques.

%Based on the above discussion, we obtain the following result for enclosing all points in $\mathbb{R}^2$ using line segments.

\begin{theorem}\label{thm:enclosing_segments}
There exists a polynomial-time $O(\alpha(n) \log n)$-approximation algorithm for \EnclosingPoints with segments, where $n$ is the total number of points and segments.
\end{theorem}

\ignore{

Our approximation algorithm also works for separating a single point with all other points.

TODO:
We should remark that despite our approximation results, it is not known whether the version of enclosing problem we study here is NP-hard. This is a question that we want to further investigate.
check NP-hardness for unit-disks set cover.

For future works on this topic, we plan to see whether our ideas can be extended to other geometric objects, and whether the approximation factor can be improved to close to constant.

Q: what is the minimum number of color classes that we can prove NP-hardness for this problem? (say 2 colors?)

A related problem is as follows:
\begin{problem}
Given $n$ points in 2D and a set of $m$ geometric objects, remove the smallest subset of objects such that all points are not enclosed (i.e.\ can go to infinity).
\end{problem}
This problem is NP-hard, via a reduction from Steiner tree on planar graphs \cite{}.

separating a single point with all other points

Q: hardness for the enclosing problem in higher dimensions, say in 3D? (our algorithm does not work)

Q: separating two sets of points $X_1$ and $X_2$.
our approx algorithm works for disks? (which has linear union complexity)

Q: another objective function: choose a number of polygons with edges being subsegments of the input segments to separate the points, minimizing the total number of turns of the polygons (i.e., minimize the number of arcs on the boundary).
%chessboard
}